\documentclass{Interspeech}

\usepackage{cite}
\usepackage{amsmath,amssymb,amsfonts}
\usepackage{algorithmic}
\usepackage{graphicx}
\usepackage{textcomp}
\usepackage{xcolor}

\usepackage{bm, amssymb, booktabs, multirow,cite, url, xcolor, hyperref}
\usepackage{makecell, arydshln}
\usepackage{threeparttable}

\usepackage[T1]{fontenc}
\usepackage[utf8]{inputenc}
\usepackage{makecell}

\interspeechcameraready

\title{Fine-tune Before Structured Pruning: Towards Compact and Accurate Self-Supervised Models for Speaker Diarization}

\author[affiliation={}]{Jiangyu}{Han}
\author[affiliation={}]{Federico}{Landini}
\author[affiliation={}]{Johan}{Rohdin}
\author[affiliation={}]{Anna}{Silnova}
\author[affiliation={}]{Mireia}{Diez}
\author[affiliation={}]{Jan}{\v{C}ernock\'{y}}
\author[affiliation={}]{Luk\'{a}\v{s}}{Burget}
\affiliation[nocounter]{Brno University of Technology}{Speech@FIT}{Czechia}
\email{\{ihan, landini, rohdin, isilnova, mireia, cernocky, burget\}@fit.vut.cz}
\keywords{speaker diarization, WavLM, fine-tuning, model compression, knowledge distillation, structured pruning}

\usepackage{comment}

\begin{document}

\maketitle

% the abstract here must exactly match the abstract entered into the paper submission system
\begin{abstract}
Self-supervised learning (SSL) models like WavLM can be effectively utilized when building speaker diarization systems but are often large and slow, limiting their use in resource-constrained scenarios. Previous studies have explored compression techniques, but usually for the price of degraded performance at high pruning ratios.
In this work, we propose to compress SSL models through structured pruning by introducing knowledge distillation. Different from the existing works, we emphasize the importance of fine-tuning SSL models before pruning.
Experiments on far-field single-channel AMI, AISHELL-4, and AliMeeting datasets show that our method can remove redundant parameters of WavLM Base+ and WavLM Large by up to 80\% without any performance degradation. After pruning, the inference speeds on a single GPU for the Base+ and Large models are 4.0 and 2.6 times faster, respectively. 
% In addition, we found that for applications with less stringent pruning ratio requirements, pre-training a large model and then pruning it to the desired size might be a better approach. However, for resource-intensive scenarios, it is more reasonable to pre-train a base model and then prune it. 
Our source code is publicly available.

\end{abstract}

\section{Introduction}

%Speaker diarization refers to the task of determining ``who spoke when'' in a multi-speaker recording. 
% From clustering-based approaches \cite{wang2018speaker, park2019auto, landini2022bayesian} to end-to-end neural diarization (EEND) \cite{fujita2019end, horiguchi20_interspeech, landini2024diaper, harkonen2024eend, chen2024attention}, the paradigm of speaker diarization has evolved in the past few years.
Recently, several speaker diarization systems \cite{tawara2024ntt, plaquet2024mambabasedsegmentationmodelspeaker, han2024leveraging} 
incorporate WavLM \cite{chen2022wavlm}, one of the state-of-the-art self-supervised learning (SSL) models, resulting in excellent performance. However, the pre-trained SSL models  often have high computational and memory costs at inference time, as well as large storage requirements, making them impractical for real-world deployment.
% To further improve performance, some diarization works \cite{tawara2024ntt, plaquet2024mambabasedsegmentationmodelspeaker, han2024leveraging} managed to incorporate WavLM \cite{chen2022wavlm}, one of the state-of-the-art self-supervised learning (SSL) models, into their frameworks. Although they got excellent results, the pre-trained SSL models usually have high costs in terms of storage, memory, and inference time, making it difficult to deploy in real scenarios. 

The challenge posed by large pre-trained models has motivated a growing interest in model compression \cite{cheng2024survey}. One widely used approach is knowledge distillation, where a small student model is trained to mimic the behavior of a large pre-trained teacher model. Prior studies \cite{chang2022distilhubert, ashihara2022deep, gandhi2023distil} have shown promising results with different student architectures. While effective, developing a good student model usually relies on expertise and might lead to suboptimal results. In contrast, structured pruning \cite{louizos2018learning, wang2019structured, xia2022structured, xia2023sheared} automatically learns a compact architecture by removing the compact sets of redundant parameters, such as attention heads and intermediate dimensions of feed-forward networks, from the large model. Compared to unstructured pruning \cite{sanh2020movement, huang2021sparse} that removes any individual parameters (i.e. individual weights in a feed-forward layer), structured pruning can lead to actual inference speed-up without support of sparse matrix operations. Besides, independently of the type of pruning, it is known that the performance can be substantially improved by introducing a distillation objective \cite{sanh2020movement, lagunas2021block, xia2022structured}. 

In the field of speech processing, current efforts mainly focus on compressing pre-trained SSL models for SUPERB \cite{yang2021superb} evaluation and speech recognition. However, existing methods often lead to significant performance drops at high pruning ratios (e.g., above 70\%, i.e. keeping less than 30\% of the original parameters) \cite{chang2022distilhubert, wang2023task, peng2023dphubert}, or only report results for low pruning ratios (e.g., under 50\%) \cite{peng2023structured, jiang2023accurate}.
% To the best of our knowledge, there is little work related to speaker diarization. 
Unlike other tasks, speaker diarization systems are usually required to handle long-form conversational recordings, where the real-world acoustic variations, conversational dynamics, and a variable number of speakers complicate the process of model compression. Therefore, efficient compression of pre-trained SSL models for diarization at high pruning ratios without sacrificing performance remains challenging and unsolved, and, to the best of our knowledge, no previous works have focused on this problem.

In this paper, we
attempt to compress the SSL models for speaker diarization while keeping the performance unchanged. 
We choose WavLM \cite{chen2022wavlm} and utilize DiariZen \cite{han2024leveraging} to build our diarization pipeline due to its competitive performance. 
For model compression, structured pruning with knowledge distillation is applied. 
Given a general SSL model, we first fine-tune the pre-trained model for the diarization task, then remove redundant parameters through structured pruning, and finally, we fine-tune the pruned model again for the diarization task. To verify the effectiveness of our method, we use the far-field single-channel data from AMI \cite{carletta2005ami, kraaij2005ami}, AISHELL-4 \cite{fu2021aishell}, and AliMeeting \cite{yu2022m2met} for system evaluation. 
Our experimental results demonstrate that our method can reduce the parameters of both WavLM Base+ and WavLM Large by up to 80\% without causing any performance loss. As a result, inference on a single GPU is sped up by 4.0 and 2.6  for the Base+ and Large models, respectively. 
% In addition, when the pruned WavLM Large has the same number of parameters as the pruned WavLM Base+, our experiments suggest that the large model tends to perform better than the base model, though excessive pruning can degrade its performance.
In addition, our experiments suggest that the Large model always outperforms the Base+ at the same level of sparsity. However,  excessive pruning (95\% for Large) can severely degrade its performance, allowing the Base+ with an equivalent number of parameters to achieve superior results.
Furthermore, we explore the relationship between the similarity of the pruned model to its original teacher and its diarization performance. Along this line, we highlight the importance of fine-tuning SSL models before pruning, which is essential to achieve reasonable results while usually overlooked by previous works \cite{wang2023task, peng2023dphubert}.

\section{Methods}
\begin{figure}[tbp]
% \vspace{-0.5cm}
  \centering
  \includegraphics[width=8cm]{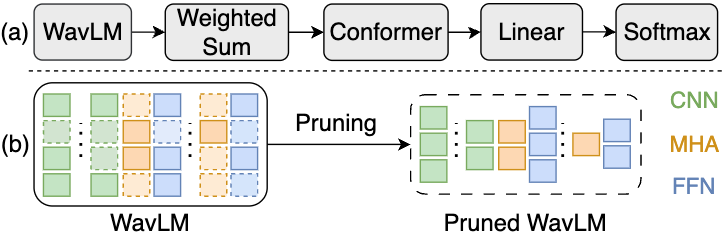}
  % \caption{ 
  % (a) overview of the whole pipeline; (b) architecture for WavLM fine-tuning; (c) illustration of structured pruning.  
  % The dashed rectangles of WavLM in (c) will be pruned.}
  \caption{ 
  (a) system for WavLM fine-tuning; (b) illustration of structured pruning.  
  The light-colored rectangles will be pruned.}
  \label{fig:framework}
  \vspace{-0.5cm}
\end{figure}

% \subsection{Overview}
% \label{subsec: overview}
% Our method consists of three steps: (1) building a diarization system on top of pre-trained WavLM and fine-tuning it for the diarization task; (2) pruning the fine-tuned model; and (3) further fine-tuning the pruned model. 
% % We focus on the application of speaker diarization with pre-trained WavLM models in this paper.
% Figure \ref{fig:framework} shows the framework of our method, where the prunable units are individual convolutional kernels in the waveform encoder convolutional neural network (CNN), entire attention heads in the multi-head attention (MHA) blocks, and the rows/columns of weight matrices corresponding to individual intermediate dimensions in the feed-forward network (FFN) blocks. 
Our method consists of three steps: (1) building a diarization system on top of pre-trained WavLM and fine-tuning it for the diarization task; (2) pruning the fine-tuned model; and (3) further fine-tuning the pruned model using the same diarization pipeline. 
% We focus on the application of speaker diarization with pre-trained WavLM models in this paper.
Figure \ref{fig:framework} shows the framework of our method, where the prunable units are individual kernels in the convolutional neural network (CNN) encoder, entire attention heads in the multi-head attention (MHA) blocks, and the rows/columns of weight matrices corresponding to individual intermediate dimensions in the feed-forward network (FFN) blocks.

\subsection{Speaker diarization pipeline}
The speaker diarization pipeline is used for WavLM fine-tuning.
We utilized DiariZen \cite{han2024leveraging}, a framework with local end-to-end neural diarization (EEND) followed by speaker embedding clustering using pyannote \cite{bredin2023pyannote}, to build our diarization pipeline.  The EEND module integrates WavLM \cite{chen2022wavlm}
with Conformer \cite{gulati2020conformer} and is trained using the powerset loss \cite{plaquet2023powerset}.
As suggested in SUPERB \cite{yang2021superb}, the WavLM outputs from each layer
are combined using weighted sum to obtain the input sequence for the following Conformer. 
We follow the original Conformer and use 4 blocks. For each block,
the input and hidden dimensions for the feed-forward module are 256 and 1024, respectively; the number of attention heads for the multiheaded self-attention is 4; for the convolution module, the kernel size is set to 31. All dropout rates in the Conformer are set to 0.1.
For powerset loss, we assume a maximum of 4 speakers and 2 overlapping speakers. 
Both WavLM Base+ and WavLM Large are explored, where the number of parameters is 94.4 million and 316.6 million, respectively.
For Conformer, the number of parameters is 6.1 million.

\subsection{Structured pruning with distillation}

As illustrated in Figure \ref{fig:framework}(b), structured pruning occurs only in WavLM. Following prior studies \cite{louizos2018learning, xia2022structured, peng2023dphubert}, we consider  optimizing  the following objective:
\begin{equation}
\label{eq:objective}
\mathcal{L}(\bm \theta) + \lambda \Vert {\boldsymbol{\theta}} \Vert_0,
\end{equation}
where $\mathcal{L}(\bm \theta)$ represents the main training loss, $\Vert {\boldsymbol{\theta}} \Vert_0$ is a regularization term that penalizes the number of non-zero parameters (thus encouraging pruning), and $\lambda > 0$ is a regularization weight controlling the level of pruning. While various choices for the main training loss are possible (see comparisons in the Section~\ref{subsec:ablation}), we achieve the best results using a distillation loss inspired by~\cite{peng2023dphubert}. Specifically, the fine-tuned WavLM is used to initialize both the fixed teacher and the student models, where the student has trainable and prunable parameters $\bm \theta$. Given the $t$-th frame, $i$-th Transformer layer outputs of the teacher $\bm h_{t}^{(i)}$ and student $\hat{\bm
 h}_t^{(i)}$, the distillation loss combines $L_1$ and cosine distances with equal weights \cite{chang2022distilhubert, peng2023dphubert}:
% \begin{equation}
%     \mathcal{L}
% %^{\text dis}
% (\bm \theta) = \sum_{i \in S} 
%     L_1(\mathbf H_{i}^{\text tea}, \mathbf H_i^{\text stu} \mathbf W_i) - cos(\mathbf H_{i}^{\text tea}, \mathbf H_i^{\text stu} \mathbf W_i)
% \end{equation}
\begin{equation}
    \mathcal{L}
(\bm \theta) = \sum_{i \in S}
\sum_{t=1}^T
    L_1(\bm h_{t}^{(i)}, \hat{\bm
 h}_t^{(i)} \mathbf W_i) - cos(\bm h_{t}^{(i)}, \hat{\bm
 h}_t^{(i)} \mathbf W_i)
\end{equation}
where $T$ is the number of frames, and $S$ denotes the set of layers used to match student to teacher via learnable linear transformations $\mathbf W_i$.
We set $S = \{0, 4, 8, 12\}$ for Base+ models and $S = \{0, 8, 16, 24\}$ for Large models. 

Unfortunately, directly optimizing \eqref{eq:objective} is intractable due to the discrete, non-differentiable nature of the $L_0$ penalty. Therefore, following \cite{louizos2018learning}, we treat ${\bm{\theta}}$ as a random variable and approximate the original loss \eqref{eq:objective} by expected value:
\begin{equation}
    \label{eq: kd_l0}
    \min_{\tilde{\bm{\theta}}, \bm{\alpha}} \mathbb{E}_{q({\bm{\theta}}|\tilde{\bm{\theta}}, \bm{\alpha})} 
    \left[ 
        \mathcal{L} 
        \left({\boldsymbol{\theta}}\right) 
        + \lambda \Vert {\boldsymbol{\theta}} \Vert_0 
    \right]
\end{equation}
where $q({\bm{\theta}}|\tilde{\bm{\theta}}, \bm{\alpha})$ is a parametric distribution of ${\bm \theta} = \{ \tilde{\bm \theta}_j z_j\}_{j=1}^J$. Here, $\tilde{\bm \theta}_j$ is the $j$-th group of prunable parameters (e.g., CNN kernels or MHA heads). Each $z_j$ is a random variable following the Hard Concrete distribution $q(z_j|\alpha_j)$ \cite{louizos2018learning}, a continuous distribution over interval $[0, 1]$ that assigns non-zero probability mass to $z_j=0$, which corresponds to pruning (zeroing out) the $j$-th group of parameters. The parameter $\alpha_j > 0$ controls this probability mass: a smaller $\alpha_j$ increases $q(z_j=0|\alpha_j)$.

As a result of optimizing \eqref{eq: kd_l0}, we obtain a distribution $q({\bm{\theta}}|\tilde{\bm{\theta}}, \bm{\alpha})$ parametrized by $\tilde{\bm \theta} = \{ \tilde{\bm \theta}_j\}_{j=1}^J$ and ${\bm \alpha} = \{ \alpha_j\}_{j=1}^J$. A lower value of $\alpha_j$ corresponds to a higher probability of pruning the $j$-th parameter group. Based on this distribution, we select a likely value of ${\bm{\theta}}$ as the parameters for the final pruned model. 

Notably, thanks to the Hard Concrete distribution, the term $\mathbb{E}_{q({\bm{\theta}}|\tilde{\bm{\theta}}, \bm{\alpha})} \left[\Vert {\boldsymbol{\theta}} \Vert_0 \right]$ can be analytically evaluated solely as a function of $\bm{\alpha}$,
%Then the regularization term in Eq. \eqref{eq: kd_l0} is calculated by: 
%\begin{equation}
%    \label{eq: l0_norm}
%    \mathbb{E}_{\mathbf{z} \sim q} \left[ \Vert {\boldsymbol{\theta}} \Vert_0 \right] = \sum_{j=1}^{n} \text{sigmoid} \left( \log \alpha_j - \beta \log \frac{-l}{r} \right)
%\end{equation}
and the gradients of $\mathbb{E}_{q({\bm{\theta}}|\tilde{\bm{\theta}}, \bm{\alpha})} \left[ \mathcal{L} \left({\boldsymbol{\theta}}\right)  \right]$ can be conveniently approximated using the reparametrization trick.
%\begin{align} 
% &v_j = \text{sigmoid}\left({\left(\log u_j - \log(1 - u_j) + \log \alpha_j \right)} /{\beta} \right), \notag \\ 
%     & z_j = \min(1, \max(0, (r - l) \cdot v_j + l))
%\end{align}
%where $u_j$ follows uniform distribution $\mathcal{U}(0, 1)$ and $\alpha_j$, $\beta$, $l < 0$ and $r > 0$ are parameter of the distribution. 
See \cite{louizos2018learning, xia2022structured} for further details.
%Following the configurations in \cite{louizos2018learning, xia2022structured, peng2023dphubert}, we set $l$, $r$, and $\beta$ to -0.1, 1.1, and 2/3, respectively.  

To precisely control sparsity (i.e., the ratio of pruned parameters to the original model size), we adopt an augmented Lagrangian formulation  \cite{wang2019structured} leading to the final loss:
\begin{align}
    \label{eq: final}
    \max_{\lambda_1, \lambda_2} \min_{\tilde{\bm \theta}, {\bm \alpha}} \mathbb{E}_{q({\bm{\theta}}|\tilde{\bm{\theta}}, \bm{\alpha})} 
    \left[ 
        \mathcal{L} 
        \left({\boldsymbol{\theta}}\right) 
%    \right] 
 + \lambda_1  (
%\mathbb{E}_{\mathbf{z} \sim q} \left[ 
\Vert {\boldsymbol{\theta}} \Vert_0
%\right]
- t) + \lambda_2 (
%\mathbb{E}_{\mathbf{z} \sim q} \left[ 
\Vert {\boldsymbol{\theta}} \Vert_0
%\right]
- t)^2 
\right] \notag
\end{align}
where $\lambda_1$, $\lambda_2 \in \mathbb R$ are learnable Lagrange multipliers,
%, the expected sparsity $\mathbb{E}_{\mathbf{z} \sim q} \left[ \Vert \tilde{\boldsymbol{\theta}} \Vert_0 \right]$ is calculated by \eqref{eq: l0_norm}, 
and $t$ is the pre-defined target sparsity.

\begin{table*}[htbp]
% \vspace{-0.cm}
  \caption{Diarization performance across different far-field datasets. Inference speedups are reported relative to the unpruned model, using a single NVIDIA RTX A5000 GPU with the input batch size optimized for maximum the GPU utilization. }
  \label{tab:overall1}
  \centering
  \begin{tabular}{l| c c c | c c c | c}
    \hline
        \multirow{2}{*}{System} & \multicolumn{3}{c|}{SSL info.} & \multicolumn{3}{c|}{collar=0s} & 
        \multirow{2}{*}{Macro} \\
        & Sparsity & \#Params. & Speedup & AMI & AISHELL-4 & AliMeeting &  \\
    \hline
    Fbank & - & - & - & 19.7 & 12.5 & 21.0 & 17.7 \\
    \hline
    \multirow{3}{*}{WavLM Base+} & 0\% & 94.4M & - & 15.6 & 11.8 &	17.7 & 15.0 \\ 
     % & 70\% & 28.4M & 3.1$\times$ & 15.8 & 11.9 & 17.7 & 15.1 \\
        & 80\% & 18.8M & 4.0$\times$ & 15.7 & 12.1 & 17.9 & 15.2 \\
        & 90\% & 9.4M & 5.7$\times$ & 17.2 & 12.1 & 19.2 & 16.1 \\
    \hline
    \multirow{3}{*}{WavLM Large} & 0\% & 316.6M & - & 14.8 & 11.3 & 16.3 & 14.1 \\ 
     % & 70\% & 94.7M & 2.1$\times$ & 15.3 & 11.2 & 16.0 & 14.2 \\
        & 80\% & 63.3M & 2.6$\times$ & 15.1 & 11.3 & 15.8 & 14.1 \\
        & 90\% & 30.6M & 3.5$\times$ & 15.7 & 11.2 & 17.6 & 14.8 \\
        % & 94\% & 18.8M & & & & & \\
        % & 97\% & 9.5M & & & & & \\
    \hline
  \end{tabular}
  \vspace{-0.15cm}
\end{table*}

\section{Experiments}
\subsection{Datasets}
We follow the data setups in \cite{han2024leveraging} to use the far-field single-channel data from AMI
\cite{carletta2005ami, kraaij2005ami}, AISHELL-4 \cite{fu2021aishell}, and AliMeeting \cite{yu2022m2met}, for system evaluation.  
Our model is trained using a combination of the three training sets, with their
corresponding development sets combined and used for validation.

\subsection{Configurations}
For fine-tuning experiments of our method, we utilize the same hyper-parameters as \cite{han2024leveraging} for model training and inference. 
The maximum number of epochs to further fine-tune the pruned model is 20.
The diarization error rate (DER) without collar is used for evaluation, and
a macro-averaged DER is reported to represent the overall performance across all datasets.

For pruning and distillation, the model is also trained using the same compound diarization training data. 
The maximum number of epochs is 30. For the first 5 epochs, target sparsity is linearly increased to the pre-defined values. Afterward, the level of sparsity remains unchanged, and the model parameters continue to update for the rest of the epochs. Following the idea in \cite{peng2023dphubert}, the pruned student model can be further distilled towards the teacher for up to 20 epochs.
Early stopping is applied if the validation loss does not decrease for 5 consecutive epochs.
The optimizer is AdamW \cite{loshchilov2018decoupled}. We set the learning rate of the main parameters $\bm \theta$ 
to 2e-4. For parameters of $\bm {\alpha}$, $\lambda_1$, and $\lambda_2$, the learning rate is 2e-2. 
Our source code is publicly available\footnote{\url{https://github.com/BUTSpeechFIT/DiariZen}}.

\subsection{Results and discussion}
\subsubsection{Overall performance}
% \begin{itemize}
%     \item Table \ref{tab:overall1}, an overview, just shows the results when sparsity is 0/80/90\%. Our method works well for both WavLM Base+ and WavLM Large. Our method achieves comparable performance to the unpruned model when removing 80\% redundant parameters. 
%     However, the speedup of WavLM Large is not as significant as that of WavLM Base+.
    
%     \item Figure \ref{fig:cnn_trans}, absolute inference time of CNN/Transformer after pruning; clearly speedup; different patterns for WavLM Base+ and WavLM Large. Params of CNN and Transformer; CNNs seem to be a bottleneck of the large model.
% \end{itemize}

% We show the pruning performance of WavLM models when sparsities are set to 0/80/90\% in Table \ref{tab:overall1}. 
In Table \ref{tab:overall1}, we show the performance of WavLM models with different levels of sparsity.
For reference, we also provide results using traditional filterbank (Fbank) features as input to the Conformer. As we can see, WavLM-based models attain much better performance than those using Fbank features. For pruning experiments of both WavLM Base+ and WavLM Large, even with 80\% of the parameters pruned, our approach delivers performance on par with the unpruned models.

Besides, we also include speedup as an additional metric to highlight the acceleration of inference after pruning.
As observed, with 80\% sparsity, the inference speed of the Base+ and Large models improves by up to 4.0 and 2.6 times, respectively, and can be further accelerated by removing 90\% of the parameters, though at
the cost of some loss in performance.

\begin{figure}[tbp]
% \vspace{-0.5cm}
  \centering
  \includegraphics[width=8cm]{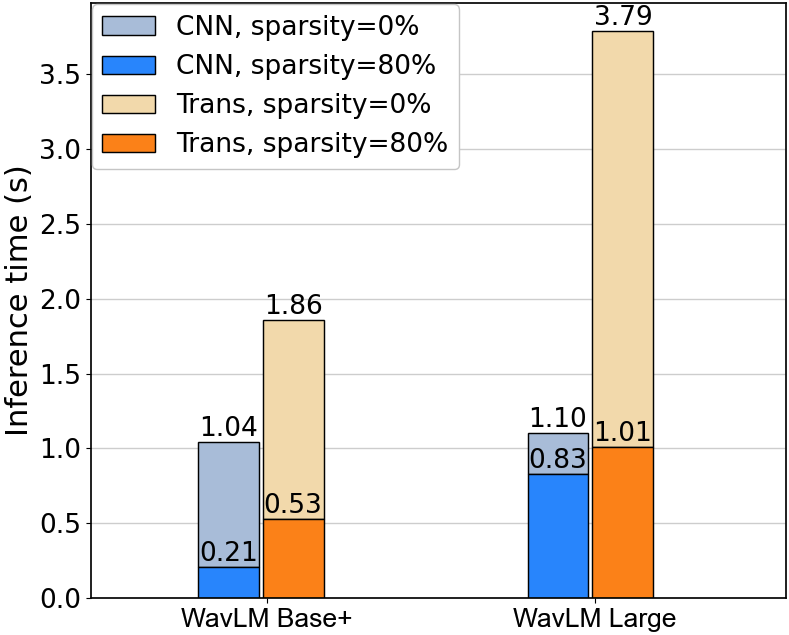}
  \caption{Inference time comparisons of CNNs and Transformer layers. The averaged values out of 5 runs are reported.}
  \label{fig:cnn_trans}
  \vspace{-0.3cm}
\end{figure}

In Figure \ref{fig:cnn_trans}, we compare the absolute inference time for CNN and Transformer layers separately and show them for both 
unpruned and pruned (80\% sparsity)
Base+ and Large models.
Note that as our aim is to fully utilize GPU, the Base+ model accepts larger batch sizes than the Large model.
As we can see, although CNNs contain a small number of parameters (4.2 million, 4.4\% for WavLM Base+ and 1.3\% for WavLM Large), they are quite resource-intensive and slow during inference, accounting for 35.9\% of the total inference time for WavLM Base+ and 22.5\% for the WavLM Large. 
After pruning,
it is clear that both models achieve notable and similar acceleration in the Transformer layers. However, the Base+ model eliminates more CNN parameters compared to the Large model, leading to a higher speedup (4.0 vs 2.6) in the end. 
Since the pruned model is learned in a fully data-driven way, 
one possible reason for the difference in CNN pruning is that the information encoded in WavLM Large is denser and more informative than in the Base+ model, making it a lower priority for pruning.

\subsubsection{Analysis under different sparsities}

% 2 plots, DER vs params; DER vs sparsity; comparisons between WavLM Base+ and WavLM Large. to see whether it's always safe to train a large model then pruning 
% 2 plots. pruning ratio of CNN and Trans to certain sparsities; WavLM Base+ and WavLM Large
As suggested in Table \ref{tab:overall1}, leveraging a pre-trained large model consistently proves advantageous. For instance, at 80\% sparsity, the pruned WavLM Large contains fewer parameters than the unpruned WavLM Base+ (63.3M vs. 94.4M) while delivering significantly better performance.
Note that both WavLM Base+ and WavLM Large are pre-trained using the same corpus, which 
raises a natural question: if our goal is to obtain a small final model, is it always beneficial to use the largest pre-trained model and then prune it to the desired size?

\begin{figure}[tbp]
\vspace{-0.1cm}   
  \centering
  \includegraphics[width=8cm]{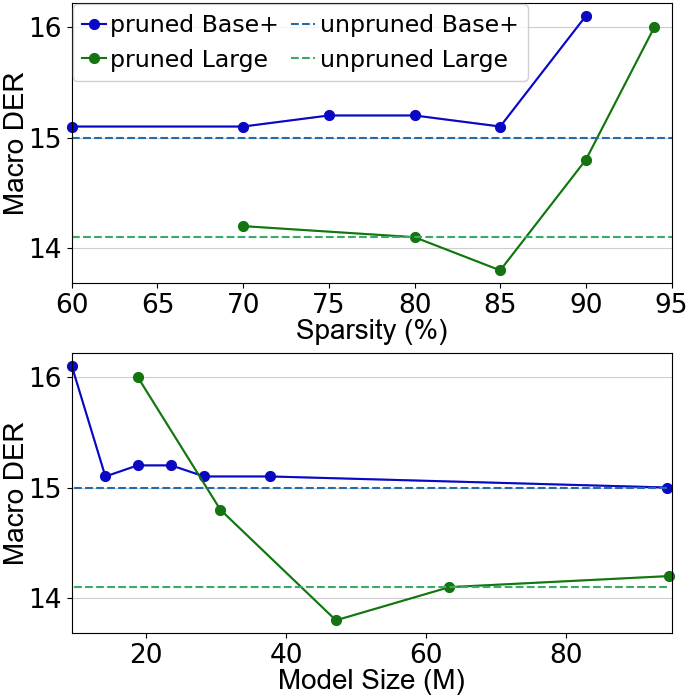}
  \caption{
  Performance comparisons between WavLM Base+ and WavLM Large under different pruning setups.}
  \label{fig:sparsity_params_base_large}
  \vspace{-0.3cm}
\end{figure}

To answer the above question, as shown in Figure \ref{fig:sparsity_params_base_large}, we explore the pruning performance of WavLM Base+ and WavLM Large under different sparsities. As indicated, for the same pruning ratio, the pruned WavLM Large models consistently outperform the pruned WavLM Base+ models. Moreover, both the Large and Base+ models can retain similar performance to the unpruned models even after removing 85\% of their parameters. However, the pruned WavLM Large at 94\% sparsity, which results in the same number of parameters (18.8M) as the pruned WavLM Base+ at 80\% sparsity, performs significantly worse. 
Our results suggest that overly aggressive pruning can completely undermine a model's performance. Thus, for scenarios requiring a fast and compact model, pruning from a smaller model may be more reasonable than pruning from a large one. 

% \vspace{-0.5mm}
\subsubsection{Ablation analysis}
\label{subsec:ablation}
% \vspace{-1mm}
% We conduct an ablation analysis of WavLM Base+ in Table \ref{tab:ablation}, including the unpruned results for reference. Initially, following previous work \cite{peng2023structured}, we adopted a combined loss of downstream task and structured pruning for model training. Specifically, the distillation part in Eq. \eqref{eq: final} was replaced with powerset loss, and the results are shown in the second line (diar + pruning).
% As observed, directly applying the prior method to our task is challenging. Moreover, fine-tuning WavLM before pruning does not help, which means that the current approach faces difficulties in yielding satisfactory results, and the well-behaved original architecture will be disrupted during pruning.

% Then we consider combining distillation and pruning objectives for model training. As we can see, fine-tuning WavLM initially is essential to maintain performance on AMI. However, the performance on AliMeeting remains significantly worse than the unpruned model. 
% This gap can be bridged by further distilling the pruned model to the original teacher.

We conduct an ablation analysis of WavLM Base+ in Table \ref{tab:ablation}. 
The unpruned results are given in the first row as a reference.
Rows 2 to 6 compare different pruning strategies, where each pruned model will be further fine-tuned to the diarization task. 
Specifically, rows 2 and 3 (diar+pruning) show results for models trained using an combination of powerset loss \cite{plaquet2023powerset} and $L_0$ norms, following the diarization pipeline in Figure \ref{fig:framework}(a). 
Rows 4 and 5 (distill+pruning) show results for pruned WavLM models trained using the distillation loss and pruning loss. The final row reports results after further distillation up to 20 epochs (without pruning) of the pruned model from the original teacher, followed by fine-tuning for diarization.
As we can see, directly applying the diarization loss and pruning loss is challenging to achieve satisfactory performance. 
When applying distillation loss during pruning, it is crucial to initialize both the teacher and student models with WavLM fine-tuned for the diarization task (preFT) to preserve unpruned performance on AMI. However, the performance on AliMeeting remains significantly worse than the unpruned model. 
This gap can be bridged by further distilling the pruned model to the original teacher.

\subsubsection{Similarity analysis}
\begin{figure}[tbp]
% \vspace{-0.5cm}
  \centering
  \includegraphics[width=8cm]{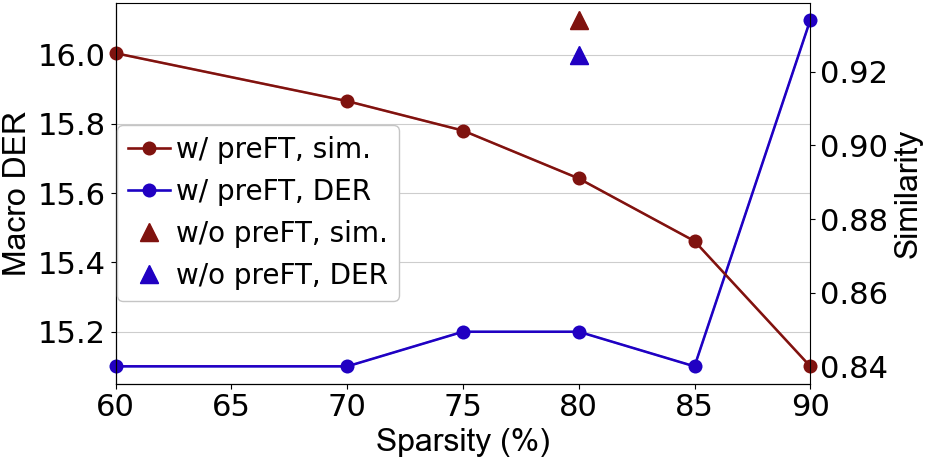}
  \caption{Macro-averaged DER and cosine similarity when using WavLM Base+ under different pruning setups. The triangle points show the performance for 80\% sparsity when distilling knowledge from the original (not fine-tuned) WavLM.}
  \label{fig:cos_sim}
  \vspace{-0.5cm}
\end{figure}

\vspace{-0.1cm}
In Figure \ref{fig:cos_sim}, we analyze the cosine similarity between the pruned model and its teacher, as well as the DER results across various pruning sparsities. The cosine similarity is averaged over layers of \{0, 4, 8, 12\} and evaluated on the whole dev data. As observed, when using the fine-tuned WavLM Base+ as a teacher, the cosine similarity decreases as sparsity increases, but it still stays fairly high even for the sparsity of 90\%. Furthermore, 
a lower cosine similarity does not result in worse performance when the sparsity is below 85\%. However, when sparsity exceeds 85\%, the cosine similarity tends to have a similar pattern to DER performance, as both start to deteriorate significantly.

When using an original WavLM Base+ as a teacher, as indicated by the triangle points, although the pruned model shows high similarity to its teacher, the DER performance is much worse. This result highlights the importance of fine-tuning before pruning. Without this, the pruned WavLM cannot be fine-tuned to achieve good performance on downstream tasks.

\subsubsection{Visualization}
\begin{figure}[tbp]
\vspace{-0.1cm}
  \centering
 \includegraphics[width=8cm]{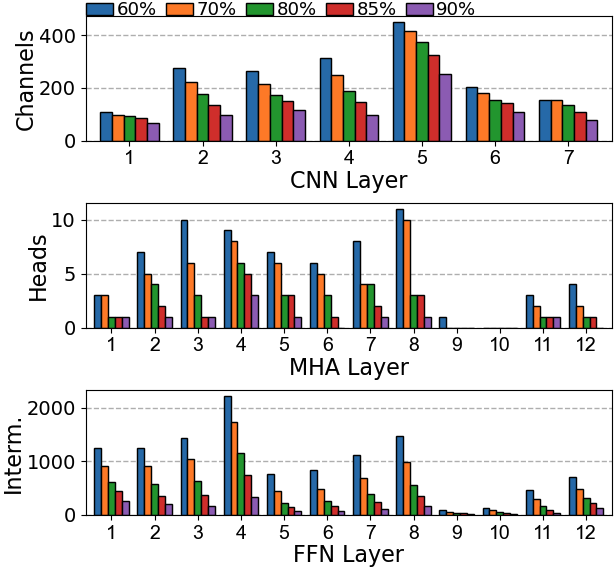}
  \caption{Visualizations of CNN channels, attention heads, and FFN intermediate dimensions from the pruned WavLM Base+. The original sizes are 512, 12, and 3072, respectively. The pruning sparsity is set to \{60\%, 70\%, 80\%, 85\%, 90\%\}.}
  \label{fig:visualization}
  \vspace{-0.5cm}
\end{figure}

\vspace{-0.1cm}
We visualize 
how the number of parameters in CNNs, MHAs, and FFNs of the pruned WavLM Base+ are changed in
Figure \ref{fig:visualization}.
% CNN channels, attention heads, and FFN intermediate dimensions of the pruned WavLM Base+ models in
% Figure \ref{fig:visualization}.
As observed, CNN layers in the middle are more relevant. For Transformer layers, the 9th and 10th are almost completely removed after pruning. For most prunable units, higher sparsity generally results in more elements being eliminated proportionally. However, for a few units, such as the 1st and last two layers of MHAs, the structure remains unchanged after a certain level of pruning, suggesting that the current structure is already dense and cannot be further pruned.
\vspace{-0.1cm}

% \begin{figure}[tbp]
% % \vspace{-0.5cm}   
%   \centering
%   \includegraphics[width=8cm]{sparsity_params_speedup_der.png}
%   \caption{
%   Effects of different pruning setups on macro-averaged DER, model size, and speedup.
%   WavLM Base+ is used.}
%   \label{fig:data_ratio}
%   % \vspace{-0.5cm}
% \end{figure}

\begin{table}[tbp]
\vspace{-0.1cm}
  \caption{Ablation analysis of pruned WavLM Base+ when sparsity is set to 80\%. preFT means fine-tuning the WavLM model to diarization task before pruning. All pruned models will be further fine-tuned to the same diarization task.}
  \label{tab:ablation}
 \setlength{\tabcolsep}{0.4mm}
  \centering
  \begin{tabular}{l| c | c c c | c}
    \hline
        \multirow{2}{*}{System} & \multirow{2}{*}{preFT} & \multicolumn{3}{c|}{collar=0s} & \multirow{2}{*}{Macro} \\
        &  & AMI & AISHELL-4 & AliMeeting &  \\
    \hline
    unpruned & - & 15.6 & 11.8 & 17.7 & 15.0 \\
    \hline
    \multirow{2}{*}{diar + pruning} & - & 16.7 & 12.0 & 18.3 & 15.7 \\
     & \checkmark & 17.0 & 12.1 & 19.7 & 16.3 \\
    \hline
    \multirow{2}{*}{distill + pruning} & - & 16.5 & 11.9 & 19.5 & 16.0 \\
     & \checkmark & 15.7 & 12.1 & 19.1 & 15.6 \\
    \quad + further distill & \checkmark & 15.7 & 12.1 & 17.9 & 15.2 \\
    \hline
  \end{tabular}
  \vspace{-0.1cm}
\end{table}

\section{Conclusion}
In this study, we propose compressing SSL models like WavLM through structured pruning, with an application to speaker diarization. 
The experiments on multiple real far-field datasets prove that our method achieves strong performance.
After pruning, our method can keep only 20\% of the parameters while maintaining the same diarization performance as with the original model. 
By doing so, an actual inference acceleration is also achieved without the need for specialized sparse matrix operations. 
% We performed a comprehensive analysis and emphasized the importance of fine-tuning before structured pruning. 
% Our method can be applied to both base and large WavLM models.
We performed a comprehensive analysis and emphasized the importance of fine-tuning pre-trained SSL models before structured pruning. Additionally, we found that pruning the largest model is not always optimal for some scenarios. 
% Our method has the potential for broader applications.
We released our code for reproducibility.

\section{Acknowledgements}
The work was supported by Ministry of Education, Youth and Sports of the Czech Republic (MoE) through the OP JAK project "Linguistics, Artificial Intelligence and Language and Speech Technologies: from Research to Applications" (ID:CZ.02.01.01/00/23\_020/0008518), by European Defence Fund project ARCHER, and Horizon 2020 Marie Sklodowska-Curie grant ESPERANTO, No.101007666. Computing on IT4I supercomputer was supported by MoE through the e-INFRA CZ (ID:90254).

\bibliographystyle{IEEEtran}
\bibliography{mybib}

\end{document}